# Pixel-wise Guidance for Utilizing Auxiliary Features in Monte Carlo Denoising

KYU BEOM HAN, Korea Advanced Institute of Science and Technology (KAIST), Republic of Korea
OLIVIA G. ODENTHAL, University of Stuttgart, Germany
WOO JAE KIM, Korea Advanced Institute of Science and Technology (KAIST), Republic of Korea
SUNG-EUI YOON, Korea Advanced Institute of Science and Technology (KAIST), Republic of Korea

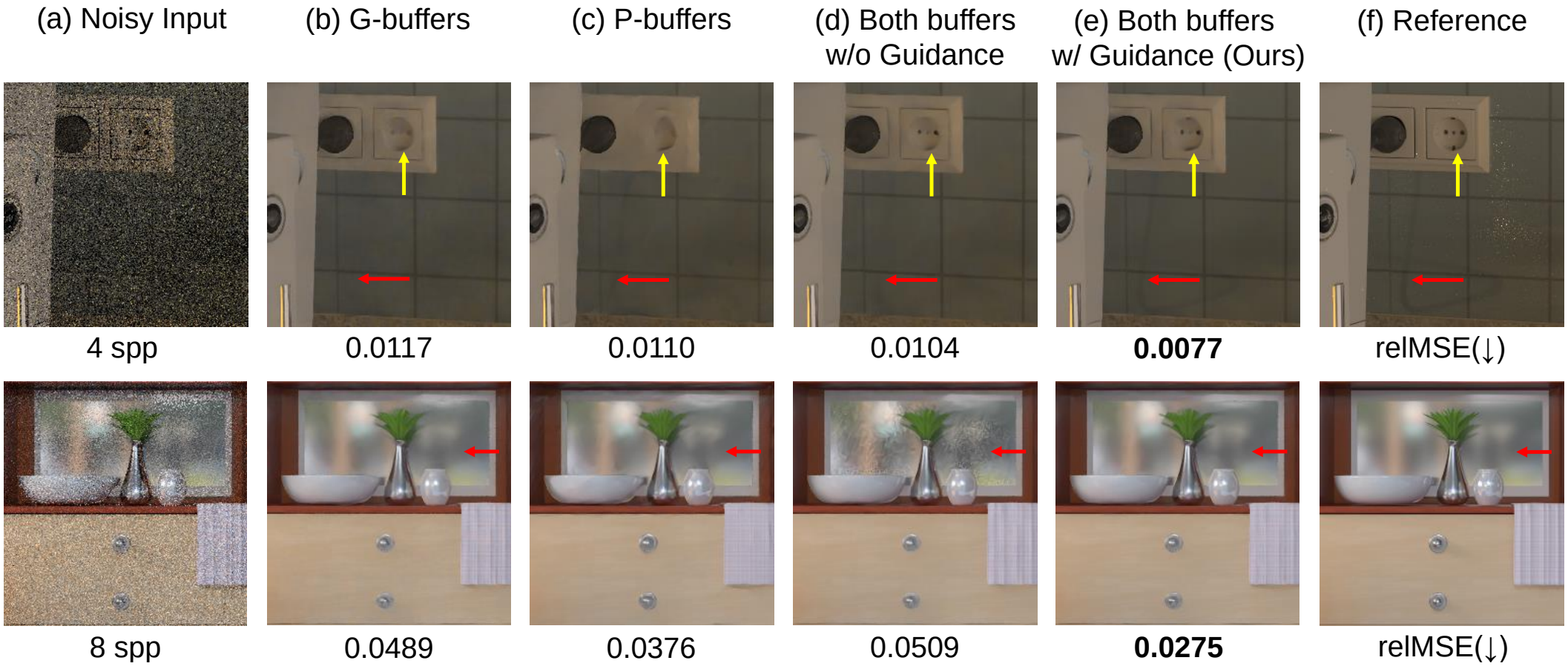

Fig. 1. We propose a novel denoising framework that provides pixel-wise guidance for utilizing auxiliary features in denoising. Denoising with geometric buffers (*G-buffers*) shows notable reconstruction in geometric details (yellow arrow of (b)), while denoising with path descriptors (*P-buffers*) better deals with complex lighting effects such as complex reflections (red arrows of (c)). However, naïve combination of *G-buffers* and *P-buffers* ignores these distinct characteristics and leads to suboptimal reconstruction of pixels with different lighting effects (d). Our framework with pixel-wise guidance effectively combines these auxiliary features to better exploit useful cues they provide (e). Best viewed with zoom-in.

Auxiliary features such as geometric buffers (*G-buffers*) and path descriptors (*P-buffers*) have been shown to significantly improve Monte Carlo (MC) denoising. However, recent approaches implicitly learn to exploit auxiliary features for denoising, which could lead to insufficient utilization of each type of auxiliary features. To overcome such an issue, we propose a denoising framework that relies on an explicit pixel-wise guidance for utilizing auxiliary features. First, we train two denoisers, each trained by a different auxiliary feature (*i.e.*, *G-buffers* or *P-buffers*). Then we design our ensembling network to obtain per-pixel ensembling weight maps,

Authors' addresses: Kyu Beom Han, Korea Advanced Institute of Science and Technology (KAIST), Republic of Korea, qbhan98@gmail.com; Olivia G. Odenthal, University of Stuttgart, Germany, olivia.odenthal@gmail.com; Woo Jae Kim, Korea Advanced Institute of Science and Technology (KAIST), Republic of Korea, wkim97@kaist.ac.kr; Sung-Eui Yoon, Korea Advanced Institute of Science and Technology (KAIST), Republic of Korea, sungeui@kaist.edu.

which represent pixel-wise guidance for which auxiliary feature should be dominant at reconstructing each individual pixel and use them to ensemble the two denoised results of our denosiers. We also propagate our pixel-wise guidance to the denoisers by jointly training the denoisers and the ensembling network, further guiding the denoisers to focus on regions where *G-buffers* or *P-buffers* are relatively important for denoising. Our result and show considerable improvement in denoising performance compared to the baseline denoising model using both *G-buffers* and *P-buffers*. The source code is available at https://github.com/qbhan/GuidanceMCDenoising.

## 1 INTRODUCTION

Generating photorealistic images for production rendering mostly relies on Monte Carlo (MC) rendering [Kajiya 1986]. MC rendering is stochastically unbiased, allowing the rendered result to eventually converge to a correct radiance. However, with short rendering times, the rendered images contain much noise and are, therefore, unsuitable for production. To this end, many denoising strategies were developed to quickly generate photorealistic images with less noise. Among these strategies, image-space MC denoising techniques [McCool 1999] are widely studied to get clean rendering results in a short time by taking a noisy image from the MC renderer and converting it to a visually clean image in a post-processing step. These denoising methods can also exploit the auxiliary features of the scene, which are easily collected during rendering.

The most popular choice for auxiliary features are geometric features (*i.e.*, *G-buffers*), which consist of albedo (texture), normal, and depth [Rousselle et al. 2013]. These features are collected during MC renderer at the non-specular first bounce of each light path. Denoising with *G-buffers* has shown robust performance on geometric details. Nevertheless, since *G-buffer* contains only the information of the first bounce of light samples, it cannot capture complex lighting effects such as multiple reflections, caustics, and participating media [Lin et al. 2021][Cho et al. 2021]. Path-space features containing information about later bounces of light samples can be used to overcome this limitation. Recent works [Gharbi et al. 2019][Lin et al. 2021][Cho et al. 2021] make use of radiometric information of each bounce of light samples, such as bounce types and sampling probability. Cho et al. [2021] further introduce a manifold learning module that effectively embeds the high-dimensional path features in low-dimensional space. These embedded features are called *P-buffer*. Generally, *P-buffers* are combined with *G-buffers* through simple concatenation as an input to various denoisers to deal with both geometric details and complex lighting effects.

These works, however, combine the distinct characteristics of these auxiliary features in an implicit manner. Each pixel of a rendered image is a result of various light effects, and the distinct effects are captured by different type of auxiliary features. For example, *G-buffers* better capture diffuse reflections and geometric details, while *P-buffers* better capture details with higher number of reflections and complex scattering. Reconstructing the details without consideration of these different characteristics could lead to suboptimal exploitation of auxiliary features as shown in the top row of Fig. 1.

To solve this problem, we propose a novel denoising method that relies on pixel-wise guidance for utilizing auxiliary feature. Our pixel-wise guidance controls how much the different auxiliary features (*i.e.*, *G-buffers* or *P-buffers*) should be exploited to reconstruct an individual pixel. To this end, we design and train independent, two-column denoisers, each of which utilizes a different type of auxiliary feature to output two different denoised images. We then design an ensembling network to learn the ensembling weight maps, which are representations of our pixel-wise guidance for utilizing auxiliary features. The ensembling weight maps are used to combine the denoised results of the denoisers. We jointly train the denoisers and the ensembling networks, guiding each denoiser to specifically focus on denoising the regions well captured by their respective auxiliary features based on the ensembling weight maps. With explicit guidance of which auxiliary feature to exploit for reconstructing certain pixels, we can better reconstruct individual pixels adaptively depending on their unique light phenomena.

We show that our design and exploitation of pixel-wise guidance on utilizing auxiliary features improves the denoising performance both numerically and perceptually. In addition, we show that our joint training scheme further improves the performance of the separate two-column denoisers. Through extensive ablation studies, we evaluate on the individual components of our framework, including the training strategies and input configurations of the ensembling network. Finally, we additionally compare our method with post-correction techniques to verify the effectiveness of our joint-training and pixel-wise guidance schemes.

## 2 RELATED WORK

### 2.1 Image-space MC Denoising and Auxiliary Features

MC rendering [Kajiya 1986] suffers from noise when using few samples per pixel (*e.g.*, 2 - 64 spp), while using a sufficient amount of samples per pixel (*e.g.*, 8000 spp) consumes much time, thus not suitable for practical applications. Image-space MC denoising techniques have been studied to address the problem by quickly removing the noise of rendered images.

To create a robust denoising model, auxiliary features for each pixel are utilized during the denoising process, where these features are easily obtained during rendering. For the classical methods, McCool et al. [1999] are the first to utilize depth and normal maps to aid their anisotropic diffusion process. Later works utilize various auxiliary features to provide an additional guide for building denoising filters [Sen and Darabi 2012][Rousselle et al. 2013][Zimmer et al. 2015][Bauszat et al. 2015]. For learning-based methods, Kalantari et al. [2015] introduce the pioneering work to apply a machine learning approach to deal with high-dimensional auxiliary features (41 channels per pixel) effectively. They train a multi-layer perceptrons (MLPs) to estimate filter weights from the auxiliary features. Later learning-based denoising works commonly use geometric features (or *G-buffers*) such as textures (*i.e.*, albedos), normals, and depths [Bako et al. 2017][Chaitanya et al. 2017][Xu et al. 2019][Yu et al. 2021], which are sufficient to provide geometric constraint for training their denoising neural networks. Furthermore, some works [Gharbi et al. 2019][Munkberg and Hasselgren 2020][Lin et al. 2021] make use of sample-wise features to utilize more fine-grain auxiliary features than aggregating them to a pixel-level.

Nevertheless, *G-buffers* only contain information on the non-specular first bounce of the light sample from the camera. This leads to misrepresentation of light samples experiencing complex lighting effects such as multiple reflections and caustics. Recent works [Gharbi et al. 2019][Lin et al. 2021][Cho et al. 2021] define path features that contain radiometric information of each bounce of light paths (*e.g.*, sampling probability and bounce type) and utilize them to deal with noises from complex lighting effects. Different types of auxiliary features are further exploited on domain-specific renderings such as volume rendering [Zhang et al. 2022] or hair rendering [Ramon Currius et al. 2022].

Due to the high dimensionality of auxiliary features, fully utilizing these features is still a challenging problem. Chaitanya et al. [2017] and Currius et al. [2022] show that the choice of auxiliary features affects the convergence of training the denoising network. Cho et al. [2021] introduce a manifold learning module that can effectively embed high-dimensional path features to a low-dimensional *P-buffers* based on a similarity in pixel radiance. On the other hand, Zhang et al. [2022] demonstrate that using all auxiliary features might degrade the denoising performance and suggest an automatic feature selection that greedily searches for the most effective subset of auxiliary features based on denoising performance and computational cost.

Our framework also tackles a similar problem of high-dimensionality by dividing the auxiliary features into two types, *G-buffers* and *P-buffers*, and denoising with each type independently. We then ensemble the results based on our predicted per-pixel weight maps. In contrast to previous work [Zhang et al. 2022] where the contribution of auxiliary features is reflected as the scalar error value of the denoised result, our weight maps provide per-pixel information of which type of auxiliary features is important in a certain image region. We argue that making use of pixel-wise importance of auxiliary features as guidance for denoising can effectively deal with pixel-wise difference in light phenomena of the image. We show the effectiveness of our per-pixel guidance and how it leads to robust denoising performance (Sec. 5.2).

### 2.2 Post-correction for MC Denoising

MC denoising methods may have systematic biases based on hand-crafted designs or training data [Zheng et al. 2021]. To remedy the problem, several works propose a post-correction method for MC denoisers by combining the pixel estimates of various denoising methods to enhance the denoising performance while reducing the bias. Recently, Zheng et al. [2021] propose an optimization-based method that searches for an effective per-pixel ensembling weight to combine multiple denoisers from various denoising methods. Based on a similar idea, some recent works aim to build a learning-based combination model that achieves unbiased denoising by combining denoised and noisy images to show consistent denoising performance on high-sample counts [Back et al. 2020][Firmino et al. 2022][Back et al. 2022][Gu et al. 2022].

Recent post-correction works focus on building a generalized combination model that works agnostic of denoiser, thus apply no modification to the denoisers. While such agnostic approach is quite useful, we explore an different approach by modifying the denoisers based on the ensembled result. We then see how much performance we can get by jointly training the denoisers along with the ensembling network to provide pixel-wise guidance of utilizing auxiliary features to the denoisers based on the per-pixel ensembling weight maps. We validate the necessity our joint-training scheme and pixel-wise guidance based on test performance (Sec. 5.1), analysis on weight maps (Sec. 5.2), and performance comparison with a post-correction method [Zheng et al. 2021] (Sec. 5.4).

## 3 PIXEL-WISE GUIDANCE OF UTILIZING AUXILIARY FEATURES

Recent denoising neural networks utilize axuiliary features in an implicit manner, resulting to insufficient exploitation of auxiliary features. To overcome the problem, we propose a novel denoising framework that denoises with an explicit pixel-wise guidance for utilizing auxiliary features. Our framework is visualized in Fig. 2. In this section, we elaborate on two stages of our work: two-column denoising (Sec. 3.1) and ensembling with per-pixel guidance (Sec. 3.2). We then demonstrate the training strategy of our denoising framework (Sec. 3.3).

### 3.1 Two-column Denoising with *G-* and *P-buffers*

In order to capture the unique nature of each type of auxiliary features and effectively utilize them for denoising, we construct two columns of denoising neural networks, $\mathcal{D}_G$ and $\mathcal{D}_P$, and separately denoise the noisy input radiances. The denoising networks $\mathcal{D}_G$ and $\mathcal{D}_P$ take the *G-buffers* and the *P-buffers*, respectively, as the input and thus utilize only a single type of auxiliary feature for

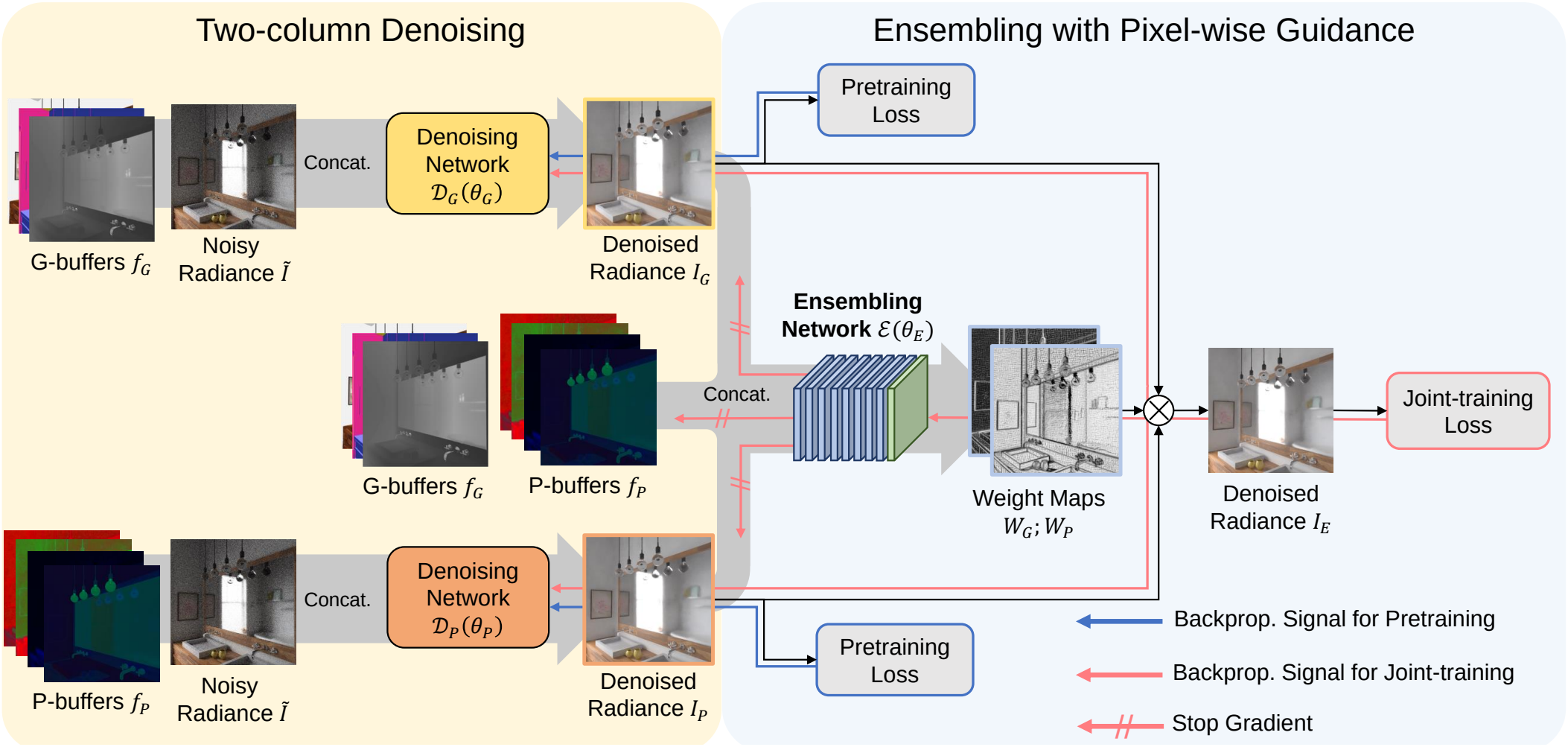

Fig. 2. Visualization of our denoising framework. First, the noisy radiance $\tilde{I}$ is fed to each column of denoisers $\mathcal{D}_G$ and $\mathcal{D}_P$ along with one type of auxiliary features (*G-buffers* $f_G$ or *P-buffers* $f_P$). The denoised results $I_G$ and $I_P$ are passed to the ensembling network $\mathcal{E}$ along with both types of auxiliary features. The ensembling network outputs per-pixel weight maps $W_G$ and $W_P$ representing the pixel-wise score of which type of auxiliary features is more effective for denoising. We apply the ensembling weights to the corresponding denoised radiance and combine them to obtain the final ensembled denoised radiance $I_E$. The path manifold module, which provides embedded *P-buffers* from path descriptors, is omitted in this figure for simplicity.

denoising. The following equations can summarize this stage:

$$I_G = \mathcal{D}_G(\theta_G; \tilde{I}, f_G), \tag{1}$$

$$I_P = \mathcal{D}_P(\theta_P; \tilde{I}, f_P), \tag{2}$$

where, $\tilde{I}$ is the input noisy image, $f_G$ and $f_P$ are *G-buffers* and *P-buffers*, respectively, and $I_G$ and $I_P$ are the denoised images of the corresponding denoising networks. $\theta_G$ and $\theta_P$ are parameters of the denoising networks $\mathcal{D}_G$ and $\mathcal{D}_P$ respectively. The two denoising networks share the same structure except for a simple modification. We change the input channel of the first layer to fit the channel dimension of the input auxiliary feature. Any architecture from the existing denoising networks can be used to construct each column of the denoising network with the simple modification to take *G-buffers* or *P-buffers* as input. We provide details of the configuration and channel number of *G-buffers* and *P-buffers* in Appendix A.

## 3.2 Ensembling with Pixel-wise Guidance

This stage aims to find appropriate per-pixel ensembling weights that will be used to provide pixel-wise guidance for our framework to utilizing auxiliary features. We design and train our ensembling network, $\mathcal{E}$, to estimate the ensembling weights. The estimated ensembling weights will be used to ensemble the denoised results $I_G$ and $I_P$ from the denoisers $\mathcal{D}_G$ and $\mathcal{D}_P$. We obtain the per-pixel weight maps, $W_G$ and $W_P$, through the ensembling network $\mathcal{E}$ as below:

$$\{W_G; W_P\} = \mathcal{E}(\theta_E; I_G, I_P, f_G, f_P), \tag{3}$$

where $\theta_E$ is the parameter of our ensembling network $\mathcal{E}$. We use *G-buffers* and *P-buffers* as inputs to the ensembling network $\mathcal{E}$ to estimate cleaner weight maps (see Sec. 5.3.2).

We construct the ensembling network based on the architecture of KPCN [Bako et al. 2017], where we modify the last convolutional layer to estimate two channels of weight maps that are passed through a softmax activation for normalization, *i.e.*, $W_G + W_P = \mathbf{1}^{H \times W}$, where $H$ and $W$ are spatial dimensions of the weight map. We build the front-end of the ensembling network with eight convolutional layers, each having fifty $5 \times 5$ convolution kernels, followed by a ReLU activation function.

The softmax activation after the last convolutional layer bounds the weight maps to $[0, 1]^{H \times W}$, making it resemble soft masks. Such normalization and range constraints are to provide stability to the ensembled result [Zheng et al. 2021]. Then we ensemble the two denoised results after applying their respective weight maps as shown below:

$$I_E = I_G \otimes W_G + I_P \otimes W_P, \tag{4}$$

where $\otimes$ is an operation that broadcasts $W$ to the dimension of $I$ and performs element-wise multiplication. The $I_E$ of Eq. 4 is the final product of our denoising framework.

Intuitively, our ensembling weight maps can be viewed as per-pixel importance map that provides pixel-wise guidance of which type of auxiliary features (*i.e.*, *G-buffers* or *P-buffers*) to use for denoising. Our ensembling network and the ensembling weight maps are critical for generating a robust ensembled result and enhancing the denoisers when jointly training the denoisers by providing pixel-wise guidance of utilizing auxiliary features. How we jointly train the ensembling network along with the denoisers is discussed in Sec. 3.3.

Our ensembling network estimates the ensembling weight maps instead of the ensembled radiance directly for stable training and fast convergence. Estimating the radiance directly is a hard task for the neural network to be trained because the radiance values are not constrained in a fixed range. This results in slower convergence when training [Bako et al. 2017]. Our weight map estimation provides the constraint by normalization through the softmax activation at the end of the ensembling network.

### 3.3 Joint-training Denoisers and Ensembling Network

We chose to jointly train our two-column denoisers and the ensembling network. Our joint-training enhances the two-column denoisers by providing pixel-wise guidance on which region each denoiser should focus on in terms of denoising with their auxiliary features. In this section, we describe the details and the effects of our joint training.

*3.3.1 Denoiser Pretraining.* Training the denoisers and the ensembling network from scratch, however, could undesirably guide the ensembling network to learn incorrect ensemble weight maps based on the denoised results from insufficiently trained denoisers. Thus, we first pretrain the two columns of denoising networks before joint-training. For pretraining the denoisers $\mathcal{D}_G$ and $\mathcal{D}_P$, we follow the training strategies of the respective denoisers we aim to adopt.

We describe the details of pretraining in Sec. 4.2.2, and provide empirical results on how pre-training leads to better convergence of the entire framework in Sec. 5.3.1.

*3.3.2 Joint-Training.* To jointly train the denoisers and the ensembling network, we design our training objective for reconstruction to minimize the loss of the final ensembled radiance $I_R$ with respect to the ground truth image $\hat{I}$. For reconstruction loss, we can use loss functions used to train the MC denoisers (*e.g.*, $\ell_1$, MSE, relative MSE [Rousselle et al. 2011], symmetric mean average percentage error (SMAPE) [Vogels et al. 2018]). We choose the loss function that the baseline denoiser use for training.

To obtain the final ensembled denoised image that is similar to the ground truth, the ensembling network learns to assign higher weights to the better-denoised result from each denoiser in a pixel-wise manner. The ensembling map thus captures the pixel-wise score on which type of auxiliary features is more effective at denoising the specific pixels.

However, the joint training based on the loss of the final ensembled result may unintentionally cause the ensembling network to also learn the denoising task. This degenerates the ensembling network to simple subsequent layers of the denoising networks, preventing the ensembling network from learning desired weight maps for ensembling. We apply stop-gradient [Chen and He 2021] on $I_G$ and $I_P$ when passing them to the ensembling network to resolve such issue.

Joint training of the denoisers and the ensembling network additionally provides the denoisers with per-pixel guidance through the ensembling weight maps. During the joint training (see back prop. arrows of Fig. 2), the ensembling weight maps $W_G$ and $W_P$ act as soft masks that preserve only the gradients corresponding to the regions they emphasize. The derivation of the gradient masking scheme is when using $\ell_1$ loss for reconstruction is shown below:

$$\frac{\partial}{\partial \theta_G} \left| I_E - \hat{I} \right| = \frac{\partial}{\partial \theta_G} \left| (I_G \otimes W_G + I_P \otimes W_P) - \hat{I} \right| \propto W_G \otimes \mathrm{sgn}(I_E - \hat{I}) \otimes \frac{\partial I_G}{\partial \theta_G}, \quad (5)$$

$$\frac{\partial}{\partial \theta_P} \left| I_E - \hat{I} \right| = \frac{\partial}{\partial \theta_P} \left| (I_G \otimes W_G + I_P \otimes W_P) - \hat{I} \right| \propto W_P \otimes \mathrm{sgn}(I_E - \hat{I}) \otimes \frac{\partial I_P}{\partial \theta_P}, \quad (6)$$

where $\mathrm{sgn}(\cdot)$ is a sign function. Note that stop-gradient makes the gradients $\frac{\partial W_P}{\partial \theta_G} = 0$ and $\frac{\partial W_G}{\partial \theta_P} = 0$.

As the weight map $W_G$ masks out the gradients for pixels with low weight scores, $\mathcal{D}_G$ is trained to reconstruct the noisy image with more emphasis on regions where *G-buffers* are more important compared to *P-buffers* for denoising. The same phenomenon happens for denoiser $\mathcal{D}_P$, which will be trained to reconstruct the noisy image with more emphasis on regions where *P-buffers* are more important compared to *G-buffers*. Therefore, our ensemble network provides guidance in which pixels each denoiser should focus on denoising with their auxiliary features. We show empirical evidence of how our joint training with pixel-wise guidance enhances the denoisers and the final ensembled result in Sec. 5.2.

## 4 EXPERIMENT SETUPS

We evaluate the performance of our denoising framework by adopting the architecture and training of KPCN [Bako et al. 2017] and comparing with the baseline denoisers. We describe our datasets, evaluation strategies, and settings for training in this section. All experiments are done on a PC with an Intel i9-10490X CPU and an NVIDIA RTX 3090 GPU.

### 4.1 Datasets and Evaluation

*4.1.1 Dataset.* We generate our dataset with Optix-based renderer [Parker et al. 2013] and scenes collected from publicly available rendering resources [Bitterli 2016] and Blend Swap. There are 18 scenes for training and validation datasets, and 32 HDRI images for the environment maps. We randomize camera parameters (*e.g.* position, up direction, look direction), material parameters (*e.g.* BRDF type, albedo, roughness, index of refraction), and environment map to create diverse training and validation datasets. However, we fixed material parameters that are crucial for physical correctness (*e.g.* specular reflectance and transmittance, interior and exterior refraction index of dielectric materials) to preserve physically correct light transport. Training with randomized dataset helps our denoising framework to deal with a wide range of light transport situations and unseen materials during test time [Yang et al. 2022].

Each rendered shot contains noisy radiance, auxiliary features required for KPCN, and path descriptors required for the path-manifold module [Cho et al. 2021]. We preprocess the rendered shot to fit the input specified for each baseline model. We construct training and validation datasets with the input rendered with 2 to 8 spp. We render a ground truth reference image with 8000 spp.

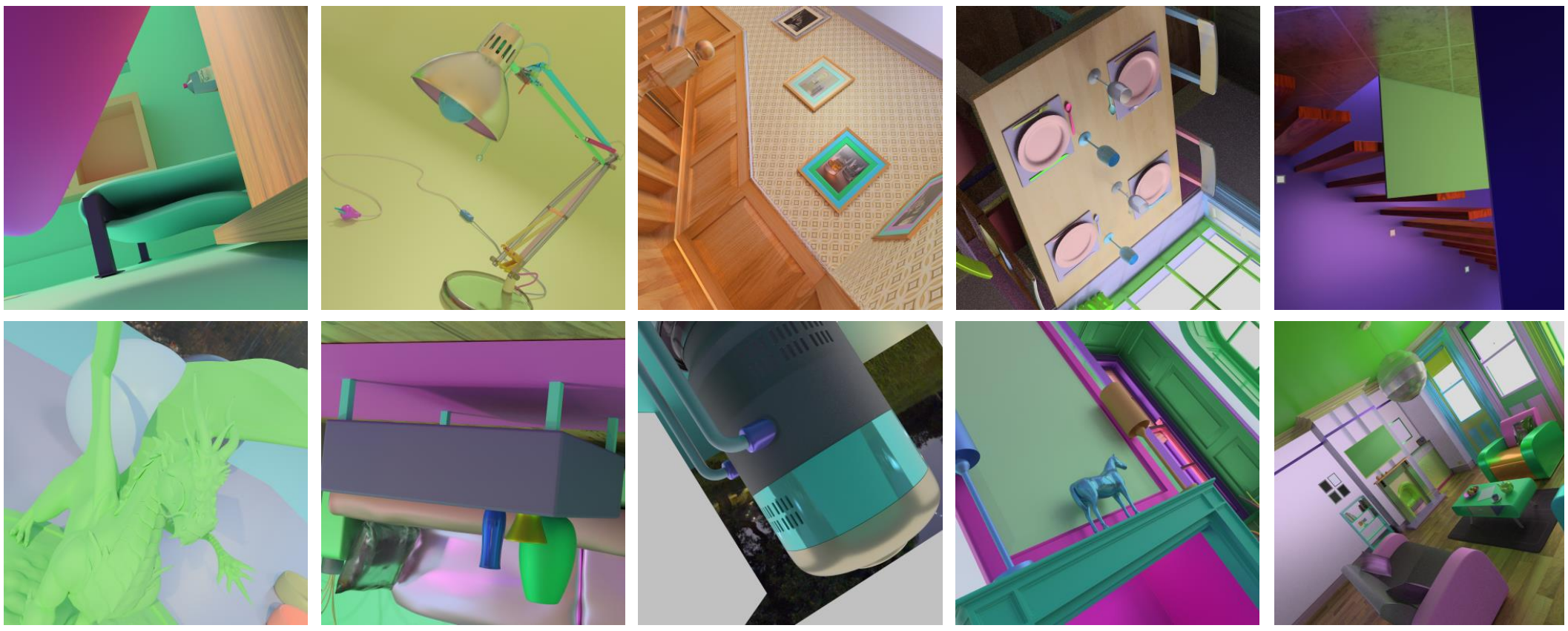

Fig. 3. Examples of clean shots from our training set. We randomize various parameters of the scene to train the denoisers with various light phenomena.

We render and preprocess 25 images per scene with the resolution of $1280 \times 1280$ for the training dataset, which makes a total of 3150 shots (see Fig. 3). For the validation dataset, we render a single shot per scene with randomization, making a total of 126 shots. During training and validation, we randomly sample 256 patches with the resolution of $128 \times 128$ from each shot on the fly and train the denoisers to denoise each patch with a batch size of 8.

Lastly, we render images for test sets with 12 scenes that are not used for training and validation. We address the configuration of *G-buffers* and *P-buffers* in Appendix A.

*4.1.2 Evaluation Metrics.* We apply gamma tone mapping [Xu et al. 2019] to the radiances for numerical and perceptual comparison on low dynamic range (LDR). To evaluate the performance on test sets, we use two error metrics: relative mean square error (relMSE) [Rousselle et al. 2011] and structural dissimilarity (1 − SSIM) [Wang et al. 2004]. We also use relMSE to evaluate the performance on the validation set. However, the error rate varies depending on the scene. To compare numerical error across the scenes, we report the errors normalized by the error of 2-spp noisy inputs for each scene and then compute the average among the scenes [Bako et al. 2017][Vogels et al. 2018][Cho et al. 2021].

## 4.2 Baseline and Training Details

*4.2.1 Baseline.* We choose two learning-based MC denoising models, each working on pixel-wise and sample-wise inputs, to show that our method is beneficial in various feature types. We use KPCN [Bako et al. 2017] as baseline for our separate denoising stage. To embed the path descriptors as *P-buffers*, we use the same path-manifold module [Cho et al. 2021].

*4.2.2 Pretraining Denoisers.* To pretrain our denoisers $\mathcal{D}_G$ and $\mathcal{D}_P$ and to train the baseline denoisers, we follow the training strategies of previous works [Bako et al. 2017][Cho et al. 2021].

For KPCN with *G-buffers*, we use $\ell_1$ loss between the reference image $\hat{I}$ and the denoised image $I$ as a reconstruction loss to train the diffuse and specular branches of KPCN. Then, we finetune both branches to minimize the $\ell_1$ loss between the reference image and the final denoised image (diffuse and specular combined).

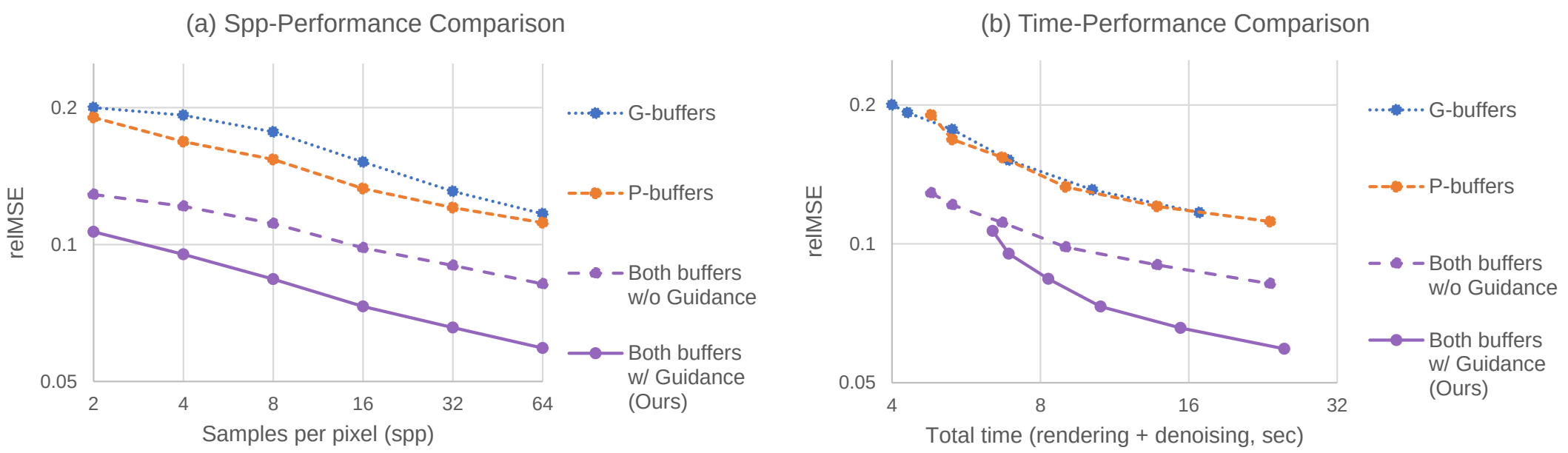

Fig. 4. We compare numerical results of our denoising framework with the baseline models using different combinations of auxiliary features. Our model outperforms their baseline models in all noise levels (left plot). Also, under the same time budget, our model shows lower error rates compared to their baseline models.

When using *P-buffers* as auxiliary features for the denoisers, we add a path-manifold module and jointly train the denoiser and the path-manifold module with the reconstruction loss and the path disentangling loss [Cho et al. 2021]. The two path-manifold modules are added to each of the diffuse and specular branches of KPCN to provide embedded *P-buffers* for each branch. The final loss term for training KPCN using *P-buffers* is as below:

$$\mathcal{L}(I, f_P, \hat{I}) = \ell_1(I, \hat{I}) + w_{path}\mathcal{L}_{path}(f_p, \hat{I}), \tag{7}$$

where $w_{path}$ is the balancing weight applied for path disentangling loss $\mathcal{L}_{path}(f_p, \hat{I})$ [Cho et al. 2021]. We use the same training objective of Eq. 7 to train KPCN using *G-buffers* and *P-buffers*, where these buffers are concatenated and fed to KPCN.

All models use ADAM optimizer [Kingma and Ba 2014] and a learning rate of $10^{-4}$ for training, and $10^{-6}$ for finetuning the KPCN. We use a *P-buffers* size of 12 and $w_{path}$ of 0.1 for KPCN which have shown the best performance for our work [Cho et al. 2021].

*4.2.3 Applying Our Framework with Joint-training.* When using KPCN for our baseline model, we construct two of our denoising frameworks for each diffuse and specular branches [Bako et al. 2017] and train each framework to minimize the $\ell_1$ loss of the ensembled result. We use a learning rate of $10^{-5}$ to train the ensembling network $\mathcal{E}$ while we use a learning rate of $10^{-6}$ to train the pretrained denoisers and the path manifold module.

Our implementation is based on a public code of baseline works, which are based on PyTorch [Paszke et al. 2019]. All convolutional layers of our model and baselines are initialized using the Xavier method [Glorot and Bengio 2010]. Pretraining and joint-training are done until the validation error of the epoch stops decreasing to prevent overfitting.

## 5 RESULTS

In this section, we first deliver the overall performance and analysis of our denoising framework by comparing it with the baseline denoisers. Then we provide ablation studies on our framework. Lastly, we report a comparison of our work with other post-correction method [Zheng et al. 2021].

### 5.1 Comparison with baseline methods

We first report the numerical comparison of our denoising frameworks and the baseline models on various noise levels (left plot of Fig. 4) and in terms of total rendering time (right plot of Fig. 4). Our framework with pixel-wise guidance achieves the best numerical result in all noise levels compared to the baseline model using *G-buffers*, *P-buffers*, and both buffers. Our framework outperforms all baseline models also in terms of time-error comparison. Even though the two-column denoising and the ensembling in our framework introduce an additional computation overhead (see Table. 1), our framework shows lower numerical denoising performance in terms of time (right plot of Fig. 4).

| Time cost (secs) v.s. spp | 2 | 4 | 8 | 16 | 32 | 64 |
|---|---|---|---|---|---|---|
| Ensembling Network $\mathcal{E}$ & Multiplication | 0.15 | 0.15 | 0.15 | 0.15 | 0.15 | 0.15 |
| $\mathcal{D}_G$ (*i.e. G-buffers*) | 1.3 | 1.3 | 1.3 | 1.3 | 1.3 | 1.3 |
| $\mathcal{D}_P$ (*i.e. P-buffers* & Both buffers w/o Guidance) | 2.1 | 2.3 | 2.7 | 3.4 | 4.9 | 7.9 |
| Both buffers w/ Guidance (Ours) | 3.7 | 3.9 | 4.3 | 5.0 | 6.5 | 9.5 |

Table 1. Runtime cost breakdown of our framework. Our framework requires two inferences of denoising networks and inference of an ensembling network, resulting in more extensive runtime than baseline models. Note that the baseline requires ensembling for each diffuse and specular branch, doubling the time cost of the ensembling. Nevertheless, our framework shows lower error rates within the same time budget compared to the baseline models thanks to our pixel-wise guidance.

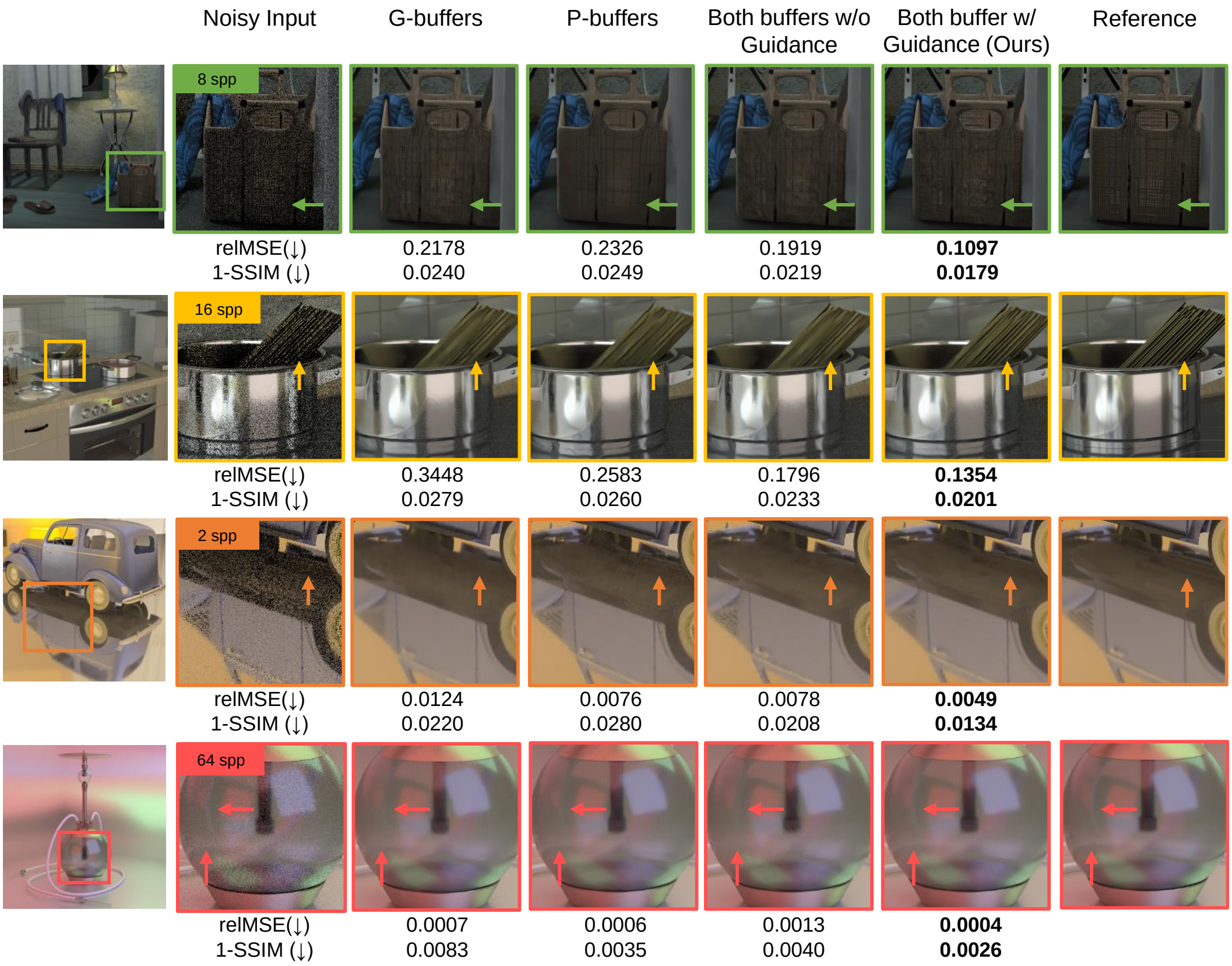

Fig. 5. Visual comparison of our denoising framework and baseline models. Our pixel-wise guidance on *G-buffers* and *P-buffers* improves the reconstruction of fine geometric details captured by *G-buffers* (first and second rows) and complex lighting effects captured by *P-buffers* (third and fourth rows). We further provide the comparison of the diffuse and specular branches in Appendix B. Best viewed with zoom-in.

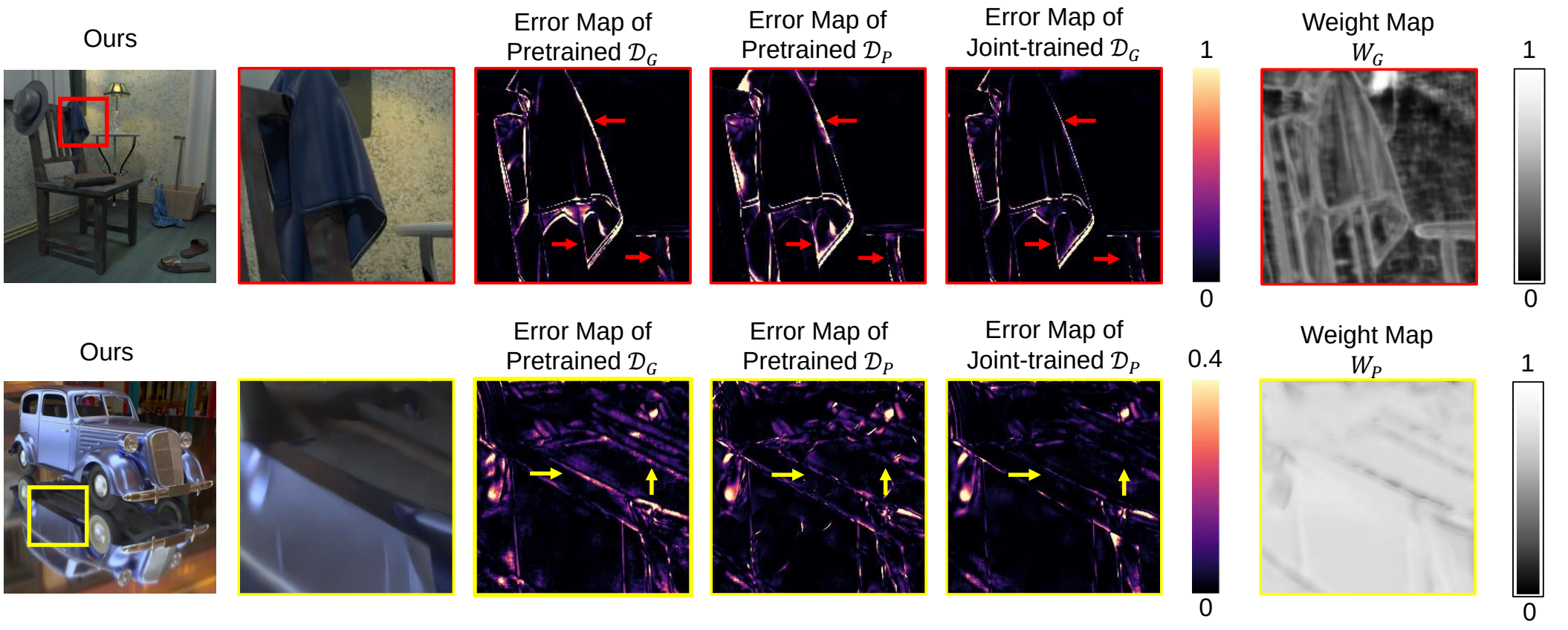

Fig. 6. Analysis on the benefits of jointly training our framework. By jointly training the denoisers with ensembling network, we enhance the denoisers to focus on regions with higher ensembling weights. This scheme reduces the error on the regions that are well reconstructed by each auxiliary feature, *G-buffers* (first row) and *P-buffers* (second row).

In Fig. 5, we provide qualitative comparison of our method and the baseline methods. Comparing the results of the first and the second rows of the Fig. 5, we can observe that our model reconstructs *G-buffer* dominant details, such as fine geometric textures, better than the baseline model using *G-buffers*. In addition, our model robustly reconstructs *P-buffer* dominant details such as reflections better than other baseline model using *P-buffers* (see third and fourth rows of Fig. 5.). Meanwhile, the baseline model using both buffers often fail to perform better than the baseline model using *G-buffers* (first row of Fig. 5), and the baseline model using *P-buffers* (third row of Fig. 5). This shows that our framework with pixel-wise guidance of utilizing auxiliary features effectively combines the advantages of *G-buffers* and *P-buffers*. We provide analysis and discussion of how our pixel-wise guidance combines the advantages of *G-buffers* and *P-buffers* in the next section.

### 5.2 Analysis on Pixel-wise Guidance for Utilizing Auxiliary Features

In this section, we analyze and discuss the importance of our pixel-wise guidance for combining the advantages of *G-buffers* and *P-buffers* in order to better utilize them for denoising. In Fig. 6, we visualize the error maps on the images denoised by each denoiser before (pretrained) and after (joint-trained) applying our pixel-wise guidance. For fair comparison, we use the same learning rate ($10^{-6}$) for joint-training the pretrained denoisers and for finetuning process during the pretraining of the denoisers to match the effect of finetuning with smaller learning rate (see Sec. 4.2.2). Our denoiser using only *G-buffers* ($\mathcal{D}_G$) shows better reconstruction on edges (third and fourth columns of first row of Fig. 6), while our denoiser using only *P-buffers* ($\mathcal{D}_P$) shows cleaner reconstruction for reflected details (third and fourth columns of second row of Fig. 6). Our pixel-wise guidance visualized by the weight maps $W_G$ and $W_P$ captures these characteristics by assigning more weights to region to focus on denoising with the given auxiliary feature (see sixth column of Fig. 6). As shown in Sec. 3.3.2, our joint-training enhances the denoisers to focus on denoising the regions with more weights. With this guidance, our framework amplifies the abilities of *G-buffers* and *P-buffers* of $\mathcal{D}_G$ and $\mathcal{D}_P$ to reconstruct such respective details, leading to decreased error on the pixels given more weights by the corresponding weight maps (fifth column of Fig. 6). These regions with lower errors further given more weight when ensembling by the weight map, resulting in a more improved ensembled result.

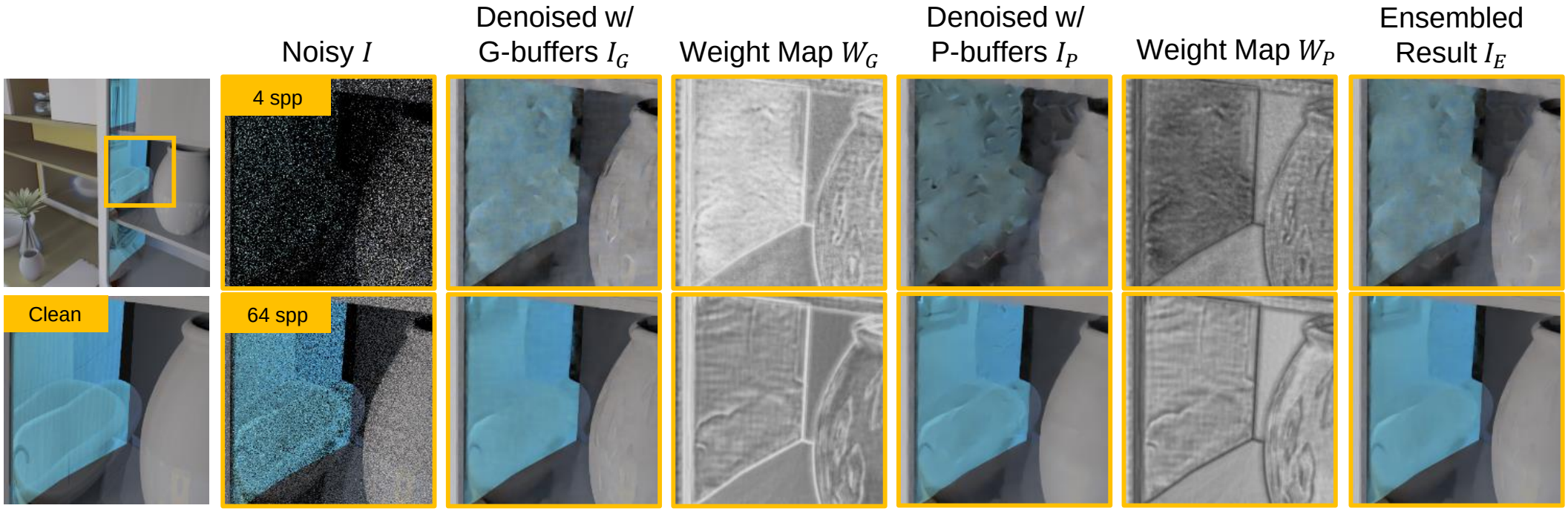

Fig. 7. Comparison of ensembling weight maps obtained for low sample counts and high sample counts. Both cases represent scenes containing a plain and glossy reflection, where *P-buffers* should play more role than *G-buffers*. However, for low sample counts (*i.e.*, 2-4 spp), denoising with *G-buffers* shows a more plausible reconstruction with fewer artifacts (third column of first row). Our ensembling network levies higher weight to results denoised with *G-buffers* in this case.

Interestingly, our novel framework can implicitly handle the noise of *P-buffers* [Cho et al. 2021], resulting in a more reliable pixel-wise guidance in different noise levels (*i.e.*, different sample counts). For the *P-buffers* to capture the full context of the light transport, the path should terminate by reaching the light source. For low sample counts, light samples are sparsely connected to the light source, adding noise to *P-buffers*. On the other hand, denoising with *G-buffers* often creates over-blurry artifacts [Back et al. 2022]. For low sample counts, however, such over-blurry reconstruction when using *G-buffers* is visually adequate than noisy reconstruction when using *P-buffers* (see third and fifth column of first row of Fig. 7). In such case, our ensembling network gives more weight to *G-buffers* and diminishes the effect of the noisy *P-buffers* even though the scene consists of reflections (see fourth column of first row of Fig. 7). On the other hand, when the samples are enough to construct reliable *P-buffers*, our pixel-wise guidance gives more emphasis to *P-buffers* than the *G-buffers* (see sixth column of second row of Fig. 7). Therefore, our pixel-wise guidance captures the noisiness of *G-buffers* and *P-buffers* on various noise levels and adaptively emphasizes more robust auxiliary features when ensembling.

## 5.3 Ablation Studies

In this section, we deliver ablation studies and the discussion in two aspects: training strategy and input configuration. The results of our ablation studies are summarized in Table 2.

*5.3.1 Training Stratagies.* We compare the test performance of our novel denoising framework trained with different strategies to show the effectiveness of our pretraining and joint-training. The numerical performances are listed in Ablation #1 of Table 2. Our training strategy (*Joint-training*) shows the lowest error rate among all noise levels compared to other training strategies. Training all models from scratch (*Full-N-Full*) shows worse performance than the baseline denoiser using both *G-buffers* and *P-buffers*. This verifies the difficulty of jointly training the denoisers and the ensembling network, reflecting the necessity of pretraining the denoisers for effective training (see Sec. 3.3.1).

Ablation #1 Training Strategies for Our Denoising Framework

| Method | Description | relMSE(↓) | 1 − SSIM(↓) |
|---|---|---|---|
| Baseline | Baseline denoiser using both buffers | 0.10529 | 0.05151 |
| *Full-N-Full* | Train all models from scratch | 0.11245 | 0.04957 |
| *Fix-N-Train* | Fix pretrained denoisers & Train ensembling model | 0.09652 | 0.04699 |
| ***Joint-training* (Ours)** | **Train pretrained denoisers & ensembling model** | **0.07983** | **0.04411** |

Ablation #2 Input Configuration for Our Ensembling Network

| Input Configuration | relMSE(↓) | 1 − SSIM(↓) |
|---|---|---|
| Denoised radiances only ($I_G$, $I_P$) | 0.10520 | 0.04846 |
| **+ *G-buffers* ($f_G$), *P-buffers* ($f_p$) (Ours)** | **0.07983** | **0.04411** |

Table 2. Ablation study on training strategies for our denoising framework (upper table) and input configuration for our ensembling network $\mathcal{E}$ (lower table).

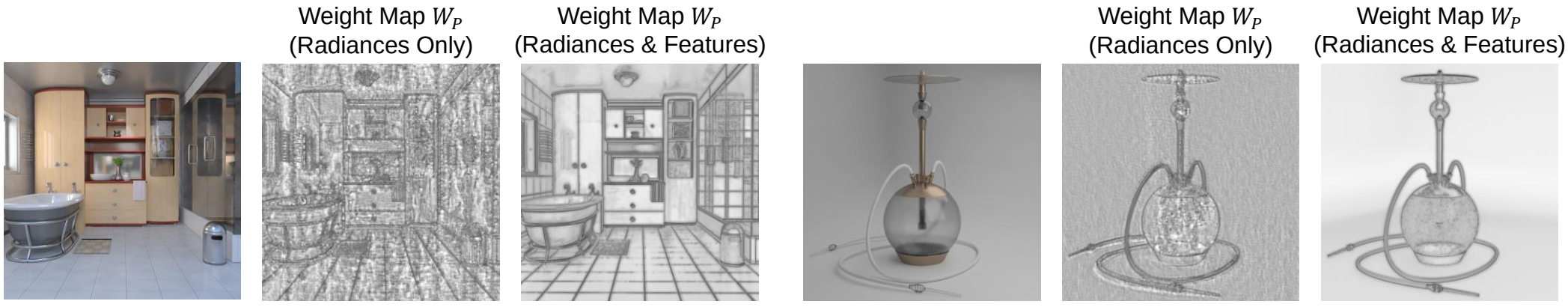

Fig. 8. Visual comparison of weight maps obtained using different input configurations. Using *G-buffers* and *P-buffers* as additional input for our ensembling network $\mathcal{E}$ returns cleaner weight maps, leading to more reliable pixel-wise guidance of utilizing auxiliary features.

*5.3.2 Input Configuration.* We additionally compare the denoising performance upon using various input configurations both numerically (Ablation #2 of Table 2) and perceptually (Fig. 8). As shown in Fig. 8, utilizing both the auxiliary features and the radiances returns much cleaner ensembling weight maps, while using only the radiances results in noisy ensembling weight maps. This is because auxiliary features provide relatively cleaner scene information compared to the radiance for the ensembling network. Using auxiliary features for cleaner emsembling weight maps shares a similar notion with Zheng et al. [2021]. They apply cross bilateral filter based on *G-buffers* to the estimated per-pixel ensembling weight maps to prevent noisy weight maps that corrupt the final ensembled result. Noisy ensembling weight maps reduce the benefit of providing rich pixel-wise guidance for denoising networks, resulting in worse test performance as also shown in Ablation #2 of Table 2.

## 5.4 Comparison with Post-correction Technique for MC Denoising

We also compare our work with one of the post-correction methods for MC denoising to verify the effectiveness of the joint-training of our framework and our pixel-wise guidance. As mentioned in Sec. 2.2, post-correction methods for MC denoising do not modify the denoisers but focus on finding appropriate ensembling weights for multiple denoisers. We chose Ensemble Denoising (ED) [Zheng et al. 2021] for comparison because this work combines the denoised result of multiple denoisers, similar to our approach. Other post-correction methods combine results of noisy (unbiased) and denoised (biased) result [Back et al. 2020][Firmino et al. 2022][Gu et al. 2022], not suitable for direct comparison with our work. Nevertheless, these methods can be applied to combine noisy image and the final ensembled image of our framework for enhanced denoising performance.

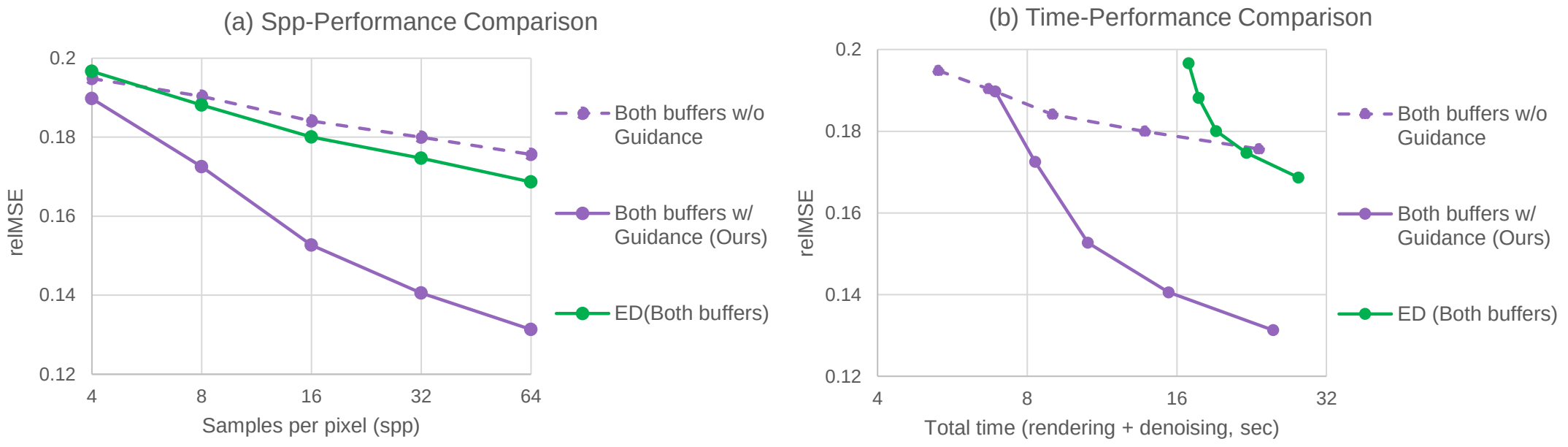

Fig. 9. Comparison of the relMSE for various spp. We compare our denoising framework with previous ensemble denoising (ED) method [Zheng et al. 2021].

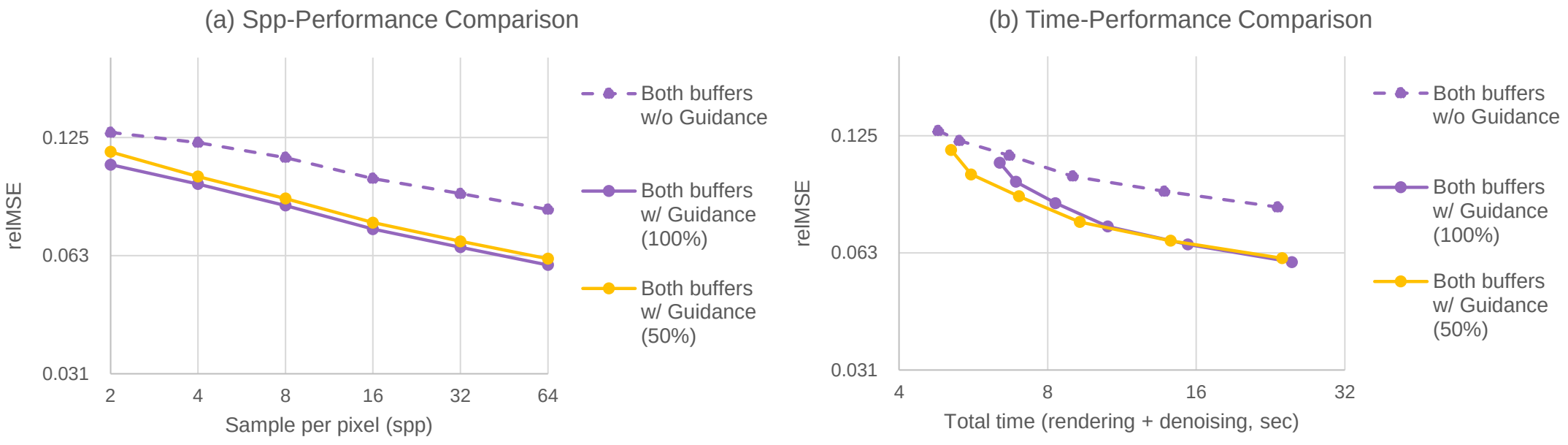

Fig. 10. Comparison of the relMSE for various spp. We compare our denoising framework with full parameters (100%) and half parameters (50%). Reducing the parameters of our framework leads to slight decrease in test performance (left plot), but becomes more comparable with the baseline using both buffers for low spp inputs time-performance efficiency (right plot).

We constructed a new test set with dual-buffer settings [Rousselle et al. 2012] to apply ED. Since our ensemble learning framework uses a per-pixel weight, we modify the implementation of ED to support per-pixel weights instead of per-color-channel weights for a fair comparison. Fig. 9 shows that our denoising framework shows better performance in all noise levels (*e.g.*, 4-64 spp) compared to ED. This reflects the effectiveness of jointly training the denoisers with the ensembling network, while the recent post-processing methods do not modify the denoisers. Moreover, our framework is more computationally efficient compared to the optimization-based ED. ED requires three denoised results from each denoiser to use their modified objective function based on a multi-buffer strategy[Rousselle et al. 2012][Back et al. 2020]. In contrast, our method requires one denoised result from each denoiser, saving time for at least one pass for each denoiser. Some learning-based post-correction methods [Back et al. 2020][Back et al. 2022] are also based on multi-buffer strategies, requiring multiple denoised results from each denoiser. Therefore, our framework has advantage over post-correction methods based on multi-buffer methods in terms of speed.

### 5.5 Analysis on Computational Cost and Performance

As mentioned in Sec. 5.1, our pixel-wise guidance framework brings a significant performance boost but with a trade-off of additional computation cost. Our framework does not show significant time-performance efficiency for low spp (*e.g.*1, 2spp) inputs due to the increased computation from two-column denoising and ensembling. To increase time-performance efficiency, we trained our framework with fewer parameters. We reduce the parameters of the denoisers and the ensembling network to 50% by reducing the number of channels to half of each convolutional layers.

The plots of numerical results of Fig. 10 show that our reduced framework shows better time-performance efficiency than our original framework. However, when given the same input, our reduced framework shows faster denoising performance, resulting in better time-performance efficiency (right plot of Fig. 10). While our original framework is 20.2-76.2% more denoising time than the baseline due to the computational cost of two-column denoising and ensembling, our reduced framework is only 6.3-14.3% slower than the baseline.

On the other hand, the denoising performance of our reduced framework is reduced for all noise level inputs (3.7-7.8%) due to the reduced number of learnable parameters (left plot of Fig. 10). Nevertheless, the performance drop of our reduced framework is less significant and still outperforms the baseline denoiser in all noise levels (left plot of Fig. 10). This reflects that the performance boost of our framework comes from the design of ensembling and joint-training, not from using more learnable parameters by adding another column of denoiser and the ensembling network.

## 6 LIMITATIONS AND FUTURE WORKS

Even though our work outperforms the baseline denoiser in the same time budget (right plot of Fig. 4 and Fig. 10), our framework is slower than a single column denoiser because our framework requires inference of two columns of denoisers as mentioned in Sec. 5.1. Our work applied to real-time MC denoisers can bring performance boost but with a lower speed.

For future works, reducing the computation of the denoising framework while maintaining meaningful pixel-wise guidance of utilizing auxiliary features will be promising. Recent approaches like generating sparse models to reduce the cost of inferencing multiple models for ensembling [Whitaker and Whitley 2022] or distilling the knowledge of a large model to a small model [Gou et al. 2021][Young et al. 2022] can maintain pixel-wise guidance while reducing the computation. In addition, searching for a light-weight but effective neural network architecture can reduce the computation. We used the architecture of KPCN [Bako et al. 2017] for implementing our ensembling network. Searching for the ideal architecture for out ensembling network can enhance the pixel-wise guidance while being cost-efficient.

Another potential direction for future works could be applying our pixel-wise guidance of auxiliary features on different denoisers and on different spaces. Ideally, since our method does not levy any constraint on model architecture, attaching our method to any type of denoiser could be promising. In this work, we have applied our method to pixel-space denoisers (*i.e.*, KPCN). Our framework is not limited to pixel-space application and could be extended to denoisers on different spaces (*e.g.*, sample-space); a potential direction for future works would be applying our pixel-wise guidance to different spaces, such as modifying our guidance as sample-wise guidance.

## 7 CONCLUSION

In this work, we have proposed a novel denoising framework that provides the denoiser with explicit pixel-wise guidance on which auxiliary features to exploit when denoising noisy pixels radiances rendered with various light phenomena. Our framework employs two denoisers trained on different types of auxiliary features, *G-buffers* and *P-buffers*, for denoising. Then, our ensembling network estimates ensembling weight maps that reflect the pixel-wise importance of each type of auxiliary features for denoising. Our ensembling weight provides pixel-wise guidance for utilizing auxiliary features by jointly training the denoisers with the ensembling network. Our framework is scalable to various types of auxiliary features by simply adding columns of denoisers, and applicable to any type of denoising networks. We believe our novel approach of pixel-wise guidance can inspire future MC denoising methods to effectively utilize their auxiliary features in various rendering domains.

## 8 ACKNOWLEDGEMENTS

We would like to thank the reviewers for their comments and suggestion for finalizing the paper. We would also like to acknowledge In-Young Cho for valuable discussions throughout the research, and the members of SGVR Lab for valuable comments on the writing. Sung-Eui Yoon is the corresponding author of the paper. This work was supported by the MSIT(Ministry of Science and ICT), Korea, under the Information Technology Research Center (ITRC) support program (IITP-2023-2020-0-01460) and National Research Foundation of Korea (NRF) (No. RS-2023-00208506).

We are grateful for the following authors and artists of the scenes used for training and testing: nacimus (Salle de bain), SlykDrako (Bedroom), cekuhnen (Coffee Maker), knightcrawler (Cornell Box, Disney Hyperion), MaTTeSr (Dining Room), Delatronic (Dragon), aXel (Glass of Water), Jay-Artist (Country Kitchen, The White Room), UP3D (Little Lamp), Wig42 (The Grey & White Room, The Chillout Room, The Modern Living Room, The Wooden Staircase, The Breakfast Room), tokabilitor (My Kitchen), piopis (Old vintage car), thecali (4060.b Spaceship, Pontiac GTO 67), NewSee2l035 (Modern Hall), Mareck (Contemporary Bathroom), NovaZeeke (Japanese Classroom), Ndakasha (BATH), Benedikt Bitterli (Material Test Ball, Utha Teapot), and ThePefDispenser (Library-Home Office).

# A CONTENTS FOR *G-BUFFERS* AND *P-BUFFERS*

## A.1 *G-buffers* Configuration for KPCN

We acquire *G-buffers* following the instructions of Bako et al. [Bako et al. 2017] as below:

- Albedo of the first non-specular bounce (3), its derivatives of $x$ and $y$ (2×3), and variance (1),
- Normal of the first non-specular bounce (3), its derivatives of $x$ and $y$ (2×3), and variance (1),
- Depth of the first non-specular bounce (1), its derivatives of $x$ and $y$ (2×1), and variance (1).

Results to a total 24 channels per pixel for each diffuse and specular branch of KPCN.

## A.2 Path Descriptor Configuration and Path Manifold Module for *P-buffers*

We follow the logic and the process of Cho et al. [Cho et al. 2021] to acquire path descriptors and embedding them to *P-buffers* using the path manifold module. Note that our OptiX-based renderer sets maximum bounce of light sample to $k = 5$. Therefore the maximum number of vertices of each sample path is 6.

The path descriptors are as below:

- Radiance undivided by the sampling probability per sample (3)
- Photon energy per sample (3)
- Sampling probability per sample (1)
- Attenuation factor of each color channel per vertex (3×6)
- Material-light interaction tag per vertex, *i.e.* reflection, transmission, diffuse, glossy, specular (1×6)
- Roughness of BSDF per vertex (1×6)

Results to a total 36 channels for path descriptors. These path descriptors are embedded as *P-buffers* via path-manifold module [Cho et al. 2021] (see Fig. 11).

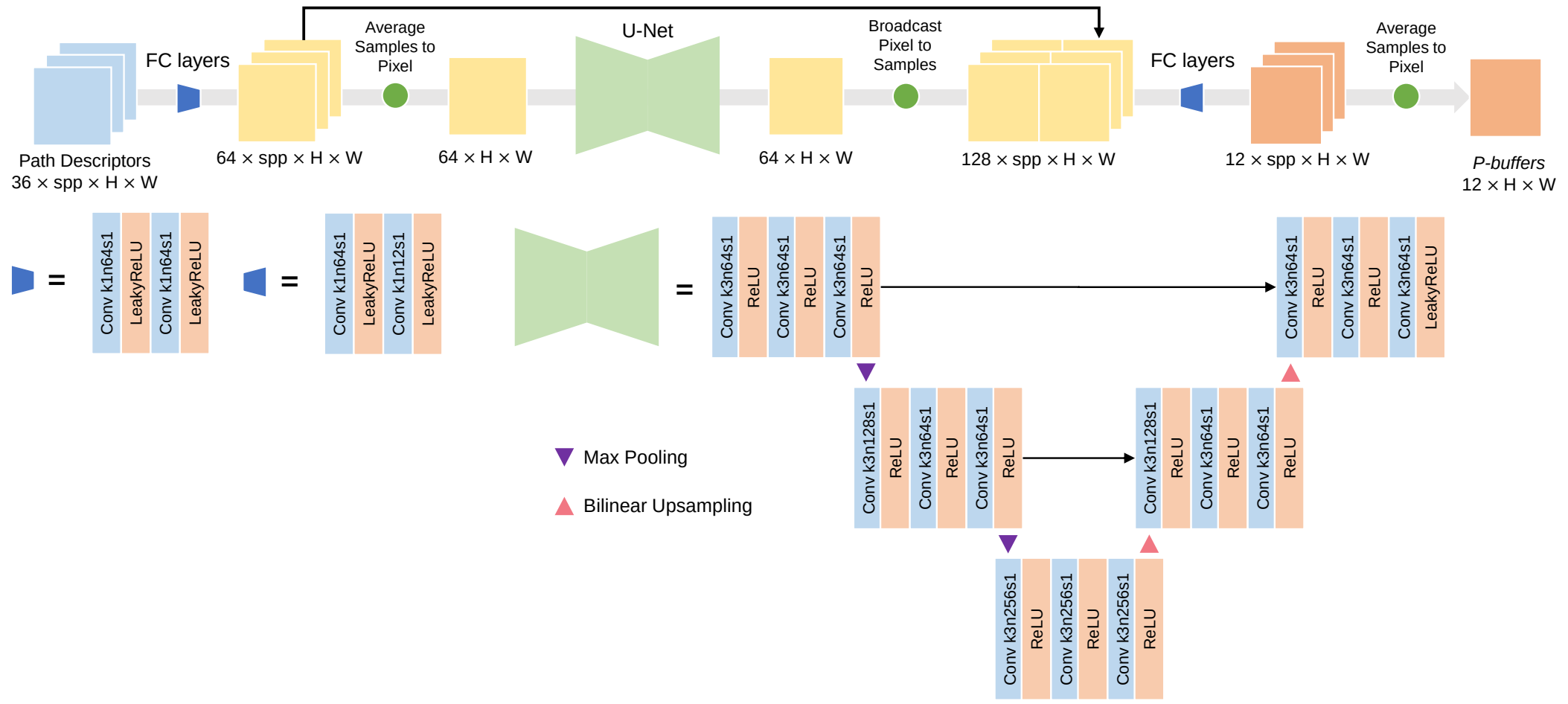

Fig. 11. Demonstration of the path-manifold module [Cho et al. 2021]. The module embeds high-dimensional path descriptors for each sample into a lower dimensional space. Then it averages the embeddings into a pixel level to be applied to the pixel-space denoiser (*e.g.* KPCN). The network is annotated in order of kernel size(k), number of channels(n) and stride(s) [Xu et al. 2019]. We add padding to preserve spatial dimension when needed, and use 0.01 as the slope for LeakyReLU.

# B QUALITATIVE RESULTS OF DIFFUSE AND SPECULAR BRANCHES

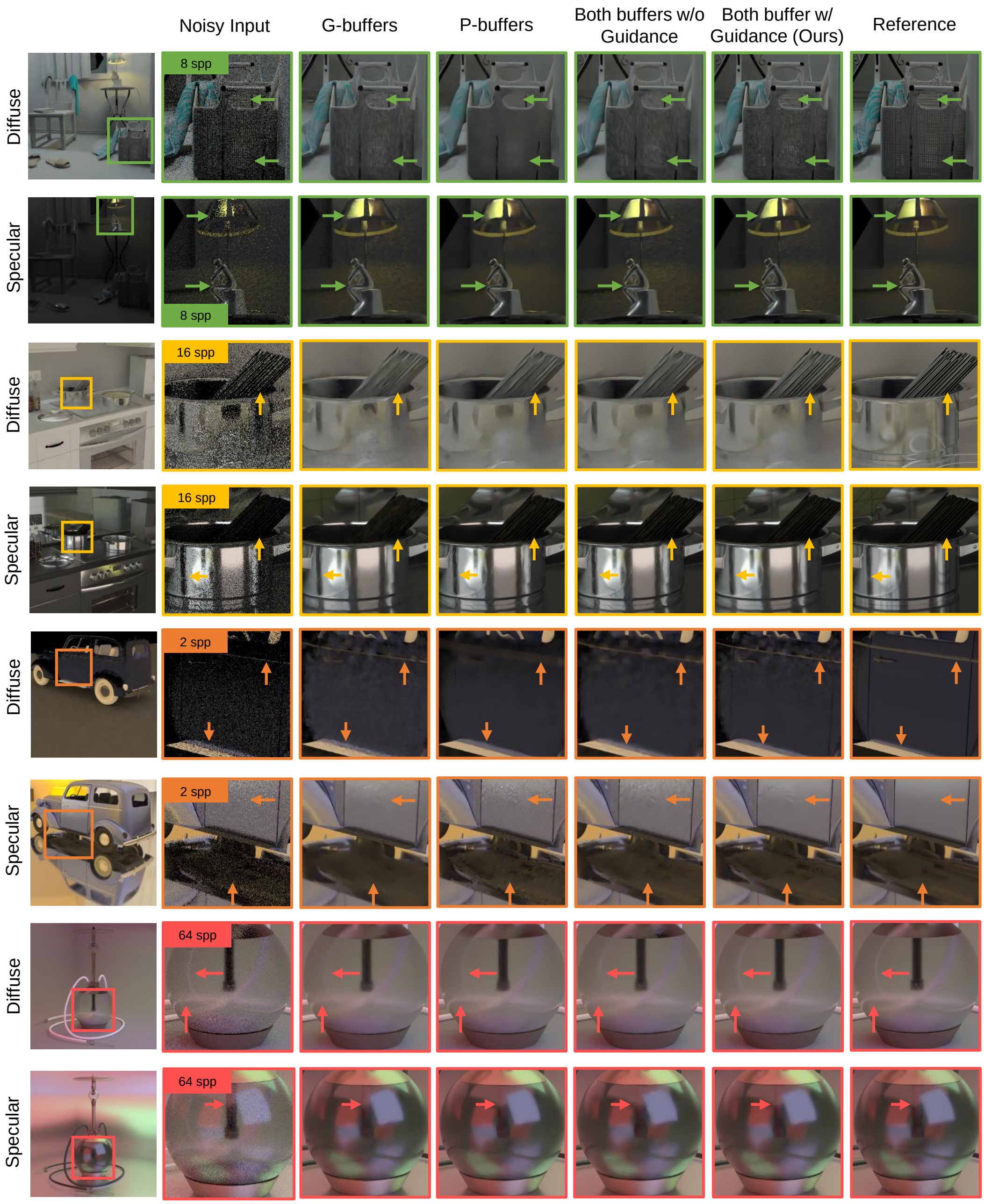

Fig. 12. Denoised results of both diffuse and specular branch of our framework and baseline models on test scenes from Fig. 5 in Fig. 12. Our results demonstrate that our pixel-wise guidance effectively exploits *G-buffers* and *P-buffers* on denoising both diffuse and specular radiances compared to the baseline.